\RequirePackage{fix-cm}
\documentclass[smallextended]{svjour3}       
\smartqed  
\usepackage{graphicx}
\begin{document}

\boldmath
\title{Nucleon Resonance Structure from Exclusive Meson Electroproduction with CLAS}
\unboldmath

\author{Victor I. Mokeev (for the CLAS Collaboration)}

\institute{Victor I. Mokeev \at
           Jefferson Laboratory, 12000 Jefferson Ave., Newport News VA, 23606, USA \\
           \email{mokeev@jlab.org}}

\date{Received: date / Accepted: date}

\maketitle

\begin{abstract}
Studies of the nucleon resonance electroexcitation amplitudes in a wide range of photon virtualities offer unique 
information on many facets of strong QCD behind the generation of all prominent excited nucleon states of distinctively 
different structure. Advances in the evaluation of resonance electroexcitation amplitudes from the data measured with 
the CLAS detector and the future extension of these studies with the CLAS12 detector at Jefferson Lab are presented. 
For the first time, analyses of $\pi^0p$, $\pi^+n$, $\eta p$, and $\pi^+\pi^-p$ electroproduction off proton channels have provided electroexcitation amplitudes of most resonances in the mass range up to 1.8 GeV and at photon virtualities 
$Q^2 < 5$~GeV$^2$. Consistent results on resonance electroexcitation amplitudes determined from different exclusive channels validate a credible extraction of these fundamental quantities. Studies of the resonance electroexcitation 
amplitudes revealed the $N^*$ structure as a complex interplay between the inner core of three dressed quarks and the  external meson-baryon cloud. The successful description of the $\Delta(1232)3/2^+$ and $N(1440)1/2^+$ electrocouplings achieved within the Dyson-Schwinger Equation approach under a traceable connection to the QCD Lagrangian and 
supported by the novel light front quark model demonstrated the relevance of dressed quarks with dynamically generated masses as an active structural component in baryons. Future experiments with the CLAS12 detector will offer insight into the structure of all prominent resonances at the highest photon virtualities, $Q^2 < 12$~GeV$^2$, ever achieved in 
exclusive reactions, thus addressing the most challenging problems of the Standard Model on the nature of hadron mass, 
quark-gluon confinement, and the emergence of nucleon resonance structures from QCD. A search for new states of 
hadronic matter, the so-called hybrid-baryons with glue as a structural component, will complete the long term efforts on 
the resonance spectrum exploration. 
\end{abstract}

\boldmath
\section{Introduction}
\unboldmath
The studies of nucleon resonances represent a particularly important avenue in the challenging adventure of exploring 
strong interaction dynamics in the regime of large quark-gluon running coupling~\cite{Bi16}. Studies of the excited 
nucleon ($N^*$) structure offer unique opportunities to explore many facets of strong QCD dynamics as it generates 
various excited nucleons with different quantum numbers of distinctively different structure~\cite{Bu17n,Ro17n}. 

Studies of the nucleon resonance electroexcitation amplitudes ($\gamma_vpN^*$ electrocouplings) within the framework 
of different quark models have demonstrated that almost all quark models are capable of reasonably well describing the nucleon elastic form factors by adjusting their parameters, but they predict a  distinctly different evolution of the $\gamma_vpN^*$ electrocouplings with photon virtualities $Q^2$ for excited nucleon states~\cite{Bu12,Bu18,Az07,Az12,Az17,Ob11,Lyub17,San15}. Comparisons 
of the experimental results on the $\gamma_vpN^*$ electrocouplings for all prominent resonances with the quark model expectations shed light on the distinctive features in the structure of excited proton states of different quantum numbers. In particular, analyses of the $\gamma_vpN^*$ electrocouplings within the quark models~\cite{Az12,Az17,Ob11} have revealed 
the structure of excited nucleons as a complex interplay between the inner core of three dressed quarks (or quark-model constituent quarks) and the external meson-baryon cloud. These findings from quark models were supported by the
conceptually different Argonne-Osaka dynamical coupled channel approach employed for a global analysis of exclusive 
meson photo-, electro-, and hadroproduction amplitudes~\cite{Suz10}.   

Particular interest to the comparative studies of the structure for the chiral-parity-partner resonance pairs was boosted by 
the outcome from the exploration of the quark distribution amplitudes (DA) for the ground-state nucleon and its chiral excited $N(1535)1/2^-$ partner. These studies were carried out by combining Lattice QCD and Light Cone Sum Rule approaches \cite{Br14,An15} constrained by the CLAS results on $N(1535)1/2^-$ electrocouplings~\cite{Az09}. They revealed pronounced differences between the quark DAs of the ground-state nucleon and its chiral partner, the $N(1535)1/2^-$ resonance. Marked differences also exist between these DAs and that of the $N(1440)1/2^+$ resonance~\cite{Me17}. In the case of chiral symmetry, relevant for the pQCD-regime, quark distributions in two chiral-parity-partners should be related just by chiral 
rotation. Furthermore, a lack of convincing evidences for the chiral parity partners in the $N^*$ spectrum in general suggests that the chiral symmetry gets broken in the strong QCD regime. A comparative study of the resonance electrocouplings for 
the chiral partner resonance pairs will allow us to explore the evolution from current to dressed quarks in the 
resonance structure and to shed light on the mechanisms of dynamical chiral symmetry breaking (DCSB), which are behind 
the generation of the dominant part of hadron mass~\cite{Ro17}.

\begin{figure}[htbp]
\centering
\includegraphics[width=8.3cm,clip]{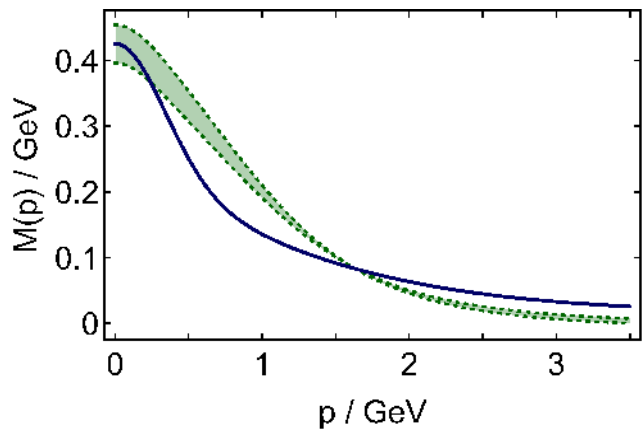}
\caption{Momentum dependence of the dressed quark dynamical mass inferred from the QCD-Lagrangian within the DSE 
approach as the solution of the gap equation tower~\cite{Bi17,Ch18} (green band) and its parameterization (solid line) with parameters fit to the data on the meson and baryon spectrum~\cite{Ch18}.}
\label{qmass} 
\end{figure}

Advances in the continuous-QCD Dyson-Schwinger equation (DSE) approach make it possible for the first time to explore the strong QCD dynamics behind the generation of the dominant part of hadron mass~\cite{Seg14,Seg15} using the results on the nucleon elastic form factors and the $Q^2$-evolution of the resonance electrocouplings. Recently, the momentum dependence of the dressed quark mass was evaluated starting from the QCD-Lagrangian within the DSE approach, as the solution of the gap equation tower, with only $\Lambda_{QCD}$ as an adjustable parameter~\cite{Bi17,Ch18}. The results are shown in Fig.~\ref{qmass}  (green band) together with the parameterization of the quark mass momentum dependence with parameters fit to the data on the meson and baryon spectra (solid line)~\cite{Ch18}. The dressed quark mass function elucidates how the almost massless current QCD-quark at momenta above 2.0~GeV becomes a fully dressed constituent-like quark of 
$\approx$400~MeV mass at momenta below 0.5~GeV.

The Higgs mechanism accounts just for mass of the current quark, relevant at large quark momenta, with the contribution less than 2\% to the fully dressed quark mass and, consequently, to the measurable hadron mass. Accounting for $<$2\% of the physical hadron mass, the Higgs mechanism is almost irrelevant in the real world of hadrons. More than 98\% of the measurable hadron mass is generated by the strong interaction in the regime of large (comparable with unity) quark-gluon running coupling. As soon as the running quark-gluon coupling becomes larger than $\approx 1/3$, a sharp increase of the dressed quark mass takes place as the manifestation of the DCSB mechanism (see Fig.~\ref{qmass})~\cite{Ro17}. Furthermore, the running quark mass also signals the emergence of quark confinement, being responsible for the violation of reflection positivity by the dressed-quark propagator, characterised by a length-scale of $\approx 0.5\,$fm~\cite{Ro16}. Mapping out the dressed quark mass function addresses the most challenging open problems of hadron physics on the nature of hadron mass and quark-gluon confinement. It makes the exploration of the dressed quark mass function using the experimental data on nucleon elastic form factors and resonance electrocouplings particularly important for the understanding of the strong QCD dynamics behind hadron generation. 

The momentum dependence of the dressed quark mass cannot be probed directly, since it is impossible to produce dressed quarks in isolation because of quark-gluon confinement. Theory input is needed in order to relate experimental results on the $Q^2$-evolution of the nucleon elastic form factors and nucleon resonance electrocouplings to the dressed quark mass function. DSE is capable of relating resonance electrocouplings to the dressed quark mass function under a traceable connection to the QCD-Lagrangian. However, relating resonance electrocouplings and dressed quark properties involves elements of modeling, which needs to be checked. Consistent results on the dressed quark mass function based on the experimental data on nucleon elastic form factors and on the electrocouplings of most of the prominent resonances with different quantum numbers and distinctively different features in their structure, analyzed independently, will validate the 
credible access to this key ingredient of strong QCD. 

The insight into strong QCD dynamics offered by the exploration of the nucleon resonance structure for all prominent excited proton states makes these studies the central direction in contemporary hadron physics.  It is an absolutely needed and important part of the efforts to achieve one of the major objective in the U.S. 2015  Nuclear Physics Long Range Plan \cite{lrp-2015} ``... using electrons to unfold the quark and gluon structure of hadrons and nuclei and to probe the Standard Model".

The advances in extracting nucleon resonance electrocouplings from the CLAS exclusive meson electroproduction data off protons will be presented in this paper, as well as their impact on exploration of strong QCD dynamics. The future extension of these studies in the upcoming experiments with the CLAS12 detector in the 12-GeV era at Jefferson Lab will be outlined.       

\boldmath
\section{Evaluation of nucleon resonance electrocouplings from exclusive meson electroproduction data with CLAS}
\unboldmath

Nucleon resonance electroexcitation can be described by the $\gamma_vpN^*$ electrocouplings $A_{1/2}(Q^2)$, $A_{3/2}(Q^2)$, and $S_{1/2}(Q^2)$ for transversely and longitudinaly polarized photons, respectively. All details on the $\gamma_vpN^*$ electrocoupling definitions and their relations to the $N \to N^*$ transition form factors can be 
found in Ref.~\cite{Bu12}. Studies of the resonance electrocoupling evolution with photon virtuality $Q^2$ offer insight into 
the internal structure of the excited nucleon states.  The $A_{1/2}(Q^2)$, $A_{3/2}(Q^2)$, and $S_{1/2}(Q^2)$ electrocouplings are extracted by fitting the experimental data on all measured observables of exclusive meson electroproduction channels 
within reaction models that are capable of reasonably well reproducing all available experimental data. Both nucleon resonances excited in the real/virtual-photon-proton $s$-channel and complex non-resonant mechanisms contribute to the full amplitude of any exclusive meson electroproduction channel. In order to access the resonance parameters, the resonant contributions to 
the full amplitudes should be isolated by employing reaction models. Currently, this is the only viable option for exclusive electroproduction processes. The credibility of the resonant contribution isolation and the resonance parameter extraction can be checked through the comparison of the resonance electrocouplings extracted independently from the data of different exclusive channels. The non-resonant contributions in different exclusive channels are entirely different. However, the resonance electrocouplings extracted from different exclusive channels should be the same, since resonance electroexcitation amplitudes cannot be affected by the resonance hadronic decays. Consistent results on the $\gamma_vpN^*$ electrocouplings obtained from independent analyses of different exclusive channels  validate the extraction of these fundamental quantities. Thus, the studies of resonance electroexcitation in different exclusive meson electroproduction channels are of particular importance for the extraction of the $\gamma_vpN^*$ electrocouplings.

Detailed studies of resonance electroexcitation in exclusive meson electroproduction off nucleons became feasible only after dedicated 
experiments were carried out with the CLAS detector~\cite{Bu17n} in Hall B at Jefferson Lab. The CLAS detector has produced the dominant part of the available world data on all meson electroproduction channels off the nucleon relevant in the resonance region for $Q^2$ up to 5.0~GeV$^2$. The results available from CLAS are summarized in Table~\ref{tab-1}. The numerical values of all observables measured with the CLAS detector are stored in the CLAS Physics Database~\cite{clasdb}. These data were obtained with almost complete coverage of the final state phase space, which is of particular importance for extraction of the resonance parameters. 

\begin{table}[htb!]
\begin{center}
\caption{\label{tab-1} Observables for exclusive meson electroproduction off protons that have been 
measured with the CLAS detector in the resonance region and stored in the CLAS Physics Database 
\cite{clasdb}: center-of-mass (CM) angular distributions for the final mesons ($\frac{d\sigma}{d\Omega}$); beam, target, and 
beam-target asymmetries ($A_{LT'}$, $A_t$, $A_{et}$); and recoil hyperon polarizations ($P'$, $P^0$).}
\begin{tabular}{|c|c|c|c|} \hline
Hadronic       &  $W$-range    & $Q^2$-range     & Measured observables \\
final state    &  GeV          & GeV$^2$         &     \\  \hline
$\pi^+ n$      &  1.10-1.38     & 0.16-0.36      & $\frac{d\sigma}{d\Omega}$ \\
               &  1.10-1.55     & 0.30-0.60      & $\frac{d\sigma}{d\Omega}$ \\
               &  1.10-1.70     & 1.70-4.50      & $\frac{d\sigma}{d\Omega}$, $A_{LT'}$ \\
               &  1.60-2.00     & 1.80-4.50      &  $\frac{d\sigma}{d\Omega}$    \\ \hline	       
$\pi^0 p$      &  1.10-1.38     & 0.16-0.36      & $\frac{d\sigma}{d\Omega}$ \\
               &  1.10-1.68     & 0.40-1.15      & $\frac{d\sigma}{d\Omega}$, $A_{LT'}$, $A_t$, $A_{et}$ \\
               &  1.10-1.39     & 3.00-6.00      & $\frac{d\sigma}{d\Omega}$  \\ \hline     
$\eta p$       &  1.50-2.30     & 0.20-3.10      & $\frac{d\sigma}{d\Omega}$ \\ \hline     
$K^+\Lambda$   &  1.62-2.60     & 1.40-3.90      & $\frac{d\sigma}{d\Omega}$ \\
               &  1.62-2.60     & 0.70-5.40      & $P'$, $P^0$ \\ \hline     
$K^+\Sigma^0$   &  1.62-2.60     & 1.40-3.90     & $\frac{d\sigma}{d\Omega}$ \\
                &  1.62-2.60     & 0.70-5.40     & $P'$ \\ \hline     
$\pi^+\pi^-p$   &  1.30-1.60     & 0.20-0.60      & Nine single-differential \\
                &  1.40-2.10     & 0.50-1.50      & cross sections \\
                &  1.40-2.00     & 2.00-5.00      &                 \\ \hline
\end{tabular}
\end{center}
\end{table}

Different approaches have been developed to extract the $\gamma_vpN^*$ electrocouplings from the measured obervables.These include the reaction models to study independently different exclusive meson electroproduction channels 
as well as global multi-channel analyses of photo-, electro-, and hadroproduction data within the coupled channel approaches. For the first time, the preliminary results on the $\Delta(1232)3/2^+$ and $N(1440)1/2^+$ electrocouplings  have become available at $Q^2$ up to 5.0~GeV$^2$ from the 8-channel global analysis of photo-, electro-, and hadro-production data. These important breakthrough results  were reported at the NSTAR2017 Conference by the Argonne-Osaka group~\cite{Ka17}. However, the results from the global multi-channel analyses are currently limited to the two lowest excited nucleon states. So far, most of the results on the $\gamma_vpN^*$ electrocouplings have been extracted from independent analyses of $\pi^+n$, $\pi^0p$, and $\pi^+\pi^-p$ exclusive electroproduction data off the proton.
  
A total of nearly 160,000 data points ($d.p.$) of unpolarized differential cross sections, longitudinally polarized beam asymmetries, and longitudinal target and beam-target asymmetries for $\pi N$ electroproduction off protons were obtained with the CLAS detector at $W < 2.0$~GeV and 0.2~GeV$^2$ $< Q^2 < 6.0$~GeV$^2$. The data have been analyzed within the framework of two conceptually different approaches: a unitary isobar model (UIM) and dispersion relations (DR)~\cite{Az09,park15,Az03}. The UIM describes the $\pi N$ electroproduction amplitudes as a superposition of $N^*$ electroexcitations in the $s$-channel and non-resonant Born terms, including $\pi$, $\rho$, and $\omega$ $t$-exchange contributions. The latter are reggeized, which allows for a better description of the data in the second- and third-resonance regions. The final-state interactions are treated as $\pi N$ rescattering in the $K$-matrix approximation~\cite{Az09,Az03}. In 
the DR approach, dispersion relations allow for the computation of the real parts of the invariant amplitudes that describe the $\pi N$ electroproduction from their imaginary parts, which are determined mostly by the resonant contributions~\cite{Az03}. 
Both approaches provide a good and consistent description of the $\pi N$ data in the range of $W < 1.7$~GeV and $Q^2 < 5.0$~GeV$^2$, resulting in $\chi^2/d.p. < 2.9$. Differences between the $\gamma_vpN^*$ electrocoupling values extracted 
by employing the UIM and DR approaches offer the systematic uncertainty estimate related to the use of the reaction models.

The $\pi^+\pi^- p$ electroproduction data from CLAS~\cite{Ri03,Fe09,Is17} provide nine independent single-differential and fully-integrated cross sections binned in $W$ and $Q^2$ in the mass range $W < 2.0$~GeV and at photon virtualities of 0.25~GeV$^2 < Q^2 < 5.0$~GeV$^2$. The analysis of these data has allowed us to develop the JM reaction model~\cite{Mo09,Mo12,Mo16} with the goal to extract resonance electrocouplings, as well as the $\pi\Delta$ and $\rho p$ hadronic decay widths. This model incorporates all relevant reaction mechanisms in the $\pi^+\pi^-p$ final-state channel that contribute significantly to the measured electroproduction cross sections off protons in the resonance region, including the 
$\pi^-\Delta^{++}$, $\pi^+\Delta^0$, $\rho^0 p$, $\pi^+N(1520)3/2^-$, $\pi^+N(1685)5/2^+$, and $\pi^-\Delta(1620)3/2^+$ meson-baryon channels, as well as the direct production of the $\pi^+\pi^-p$ final state without formation of intermediate unstable hadrons. The contributions from well established $N^*$ states in the mass range up to 2.0~GeV were included into 
the amplitudes of the $\pi\Delta$ and $\rho p$ meson-baryon channels by employing a unitarized version of the Breit-Wigner ansatz~\cite{Mo12}. The JM model provides a good description of the $\pi^+\pi^- p$ differential cross sections at $W < 2.0$~GeV and 0.2~GeV$^2 < Q^2 < 5.0$~GeV$^2$ with $\chi^2/d.p. < 3.0$ accounting for only the statistical uncertainties of
the data. The quality of the description of the CLAS data suggests the unambiguous and credible separation between the resonant and non-resonant contributions achieved by fitting the CLAS data~\cite{Mo16}. The credible isolation of the resonant contributions makes it possible to determine the resonance electrocouplings along with the $\pi \Delta$ and $\rho N$ decay widths by employing for the description of their amplitudes the unitarized Breit-Wigner ansatz~\cite{Mo12} that fully accounts for the unitarity restrictions on the resonant amplitudes. 
 
\begin{table}[htb!]
\begin{center}
\caption{\label{tab-2} Resonance electrocouplings available from the analyses of the CLAS data on exclusive meson electroproduction off protons in the resonance region.}
\begin{tabular}{|c|c|c|} \hline
Exclusive        &  Excited proton    & Coverage over $Q^2$ for extracted    \\
channel          &  state             & $\gamma_{v}pN^*$ electrocouplings, GeV$^2$    \\ \hline
$\pi^+ n$, $\pi^0 p$       &  $\Delta(1232)3/2^+$,                                & 0.16\textemdash 6.00       \\
                           &  $N(1440)1/2^+$, $N(1520)3/2^-$, $N(1535)1/2^-$      & 0.30\textemdash 4.16       \\ \hline	       
$\pi^+ n$      &  $N(1675)5/2^-$, $N(1680)5/2^+$, $N(1710)1/2^+$                  & 1.60\textemdash 4.50        \\ \hline     
$\eta p$       &  $N(1535)1/2^-$        & 0.20\textemdash 2.90       \\ \hline     
$\pi^+\pi^-p$   & $N(1440)1/2^+$, $N(1520)3/2^-$       & 0.25\textemdash  1.50       \\
                &  $\Delta(1620)1/2^-$, $N(1650)1/2^-$, $N(1680)5/2^+$      & 0.50\textemdash 1.50       \\
                &  $\Delta(1700)3/2^-$, $N(1720)3/2^+$, $N'(1720)3/2^+$     &         \\ \hline
\end{tabular}
\end{center}
\end{table}

\begin{figure}[htb!]
\centering
\includegraphics[width=3.9cm,clip]{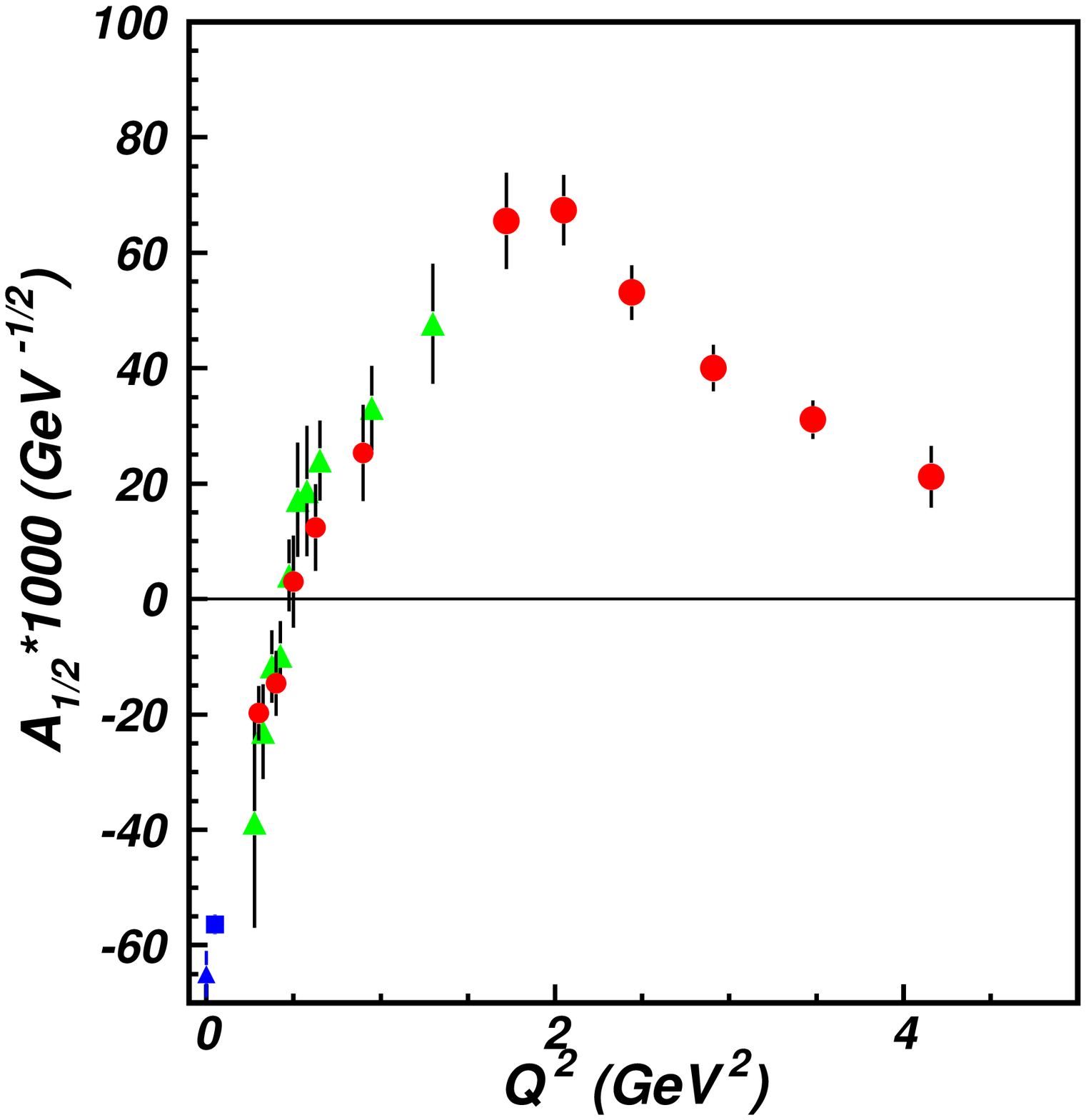}
\includegraphics[width=3.9cm,clip]{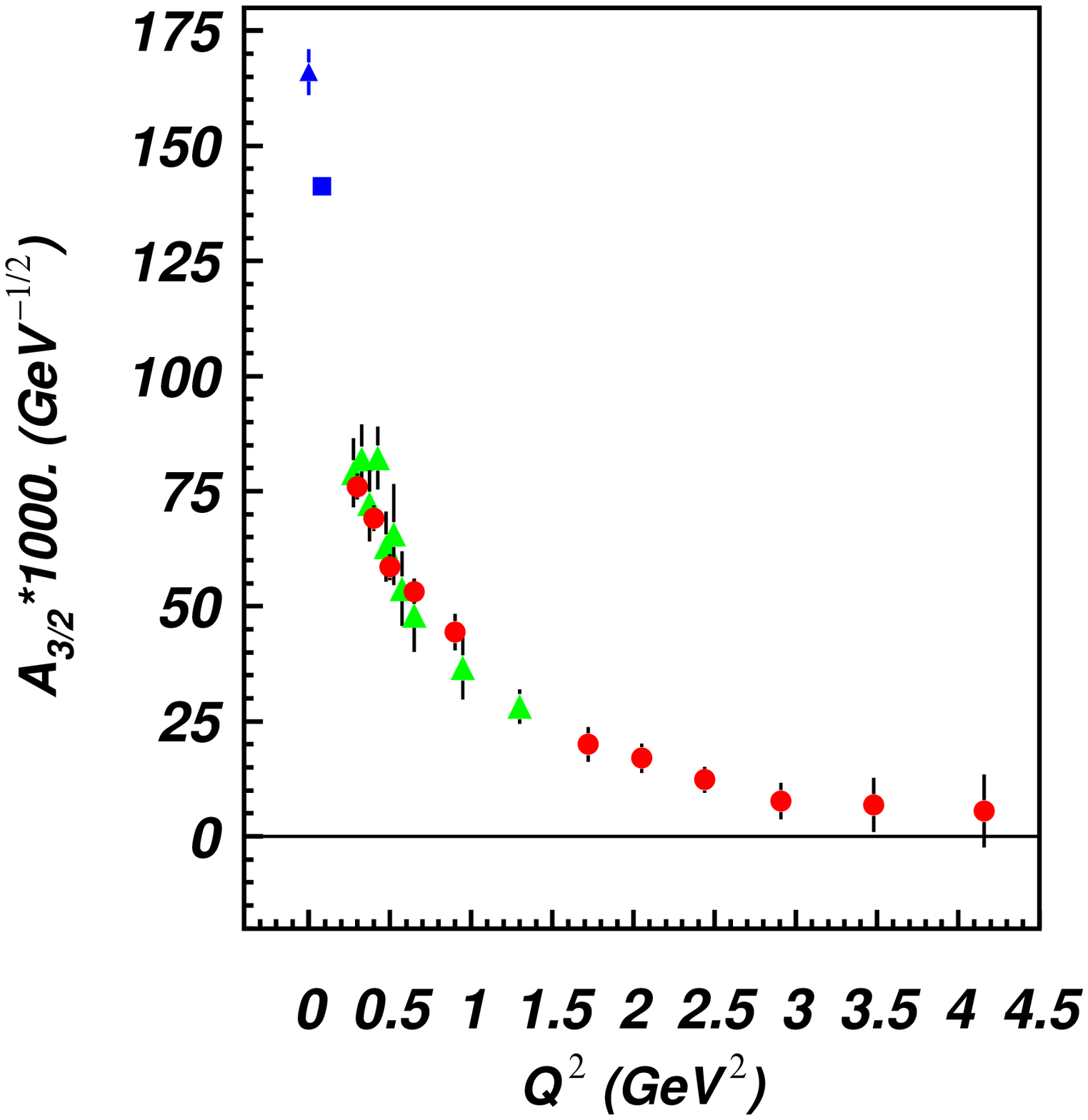}
\includegraphics[width=3.9cm,clip]{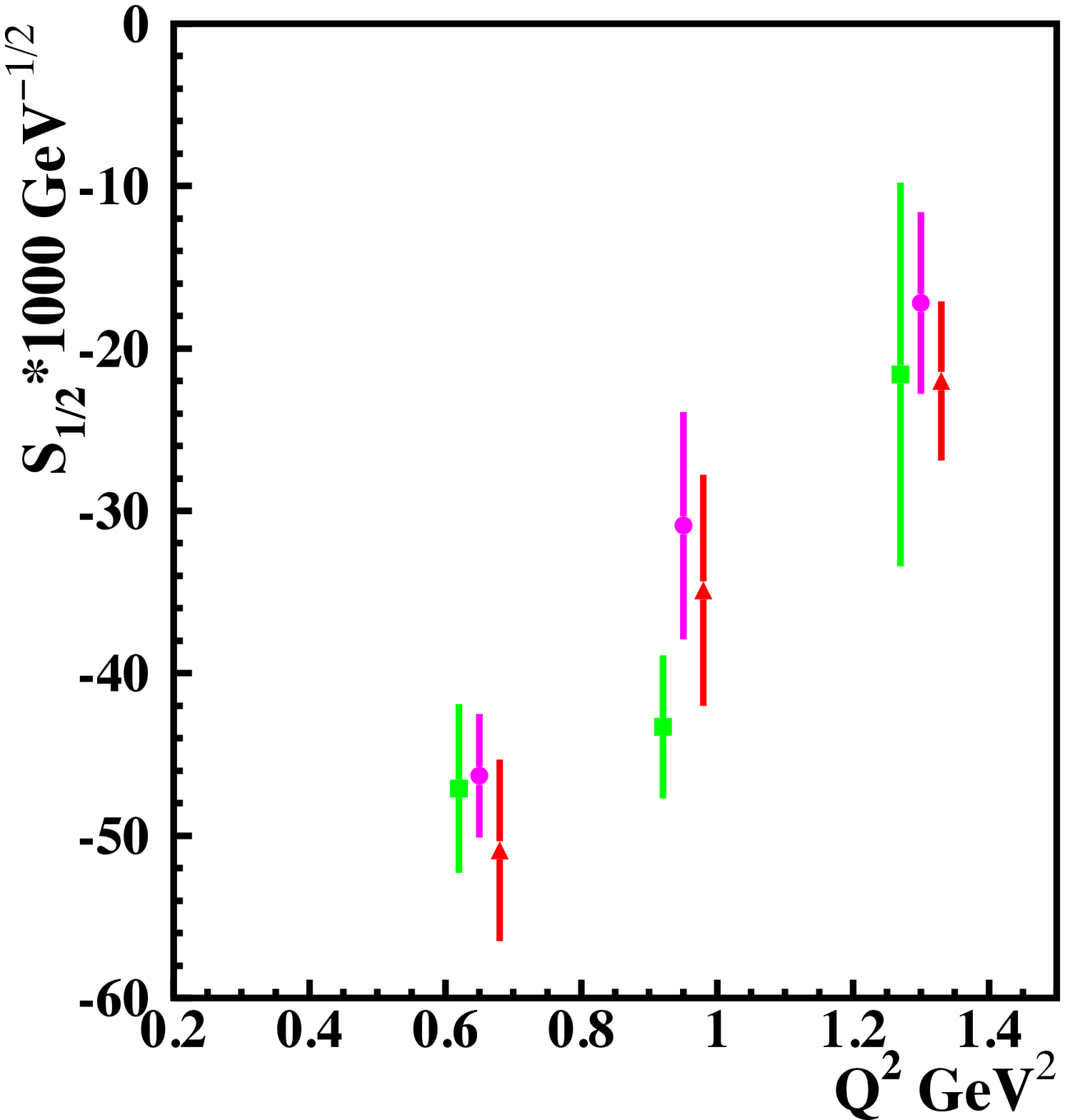} 
\caption{(Color Online) $A_{1/2}$ $\gamma_vpN^*$ electrocouplings of the $N(1440)1/2^+$ (left), $A_{3/2}$ 
$\gamma_vpN^*$ electrocouplings of the $N(1520)3/2^-$ (middle), and  $S_{1/2}$ $\gamma_vpN^*$ electrocouplings of the $\Delta(1620)1/2^-$ (right) from analyses of the CLAS electroproduction data off protons in the $\pi N$~\cite{Az09,park15} 
(red circles in the left and middle panels) and $\pi^+\pi^-p$ channels~\cite{Mo12,Mo16} (green triangles in the left and middle 
panels). The right panel shows the $\Delta(1620)1/2^-$ electrocouplings obtained from analyses of $\pi^+\pi^-p$ electroproduction data off protons \cite{Mo16} carried out independently in three intervals of $W$: 1.51~GeV $\to$ 1.61~GeV (green squares), 1.56~GeV $\to$ 1.66~GeV (magenta circles), and 1.61~GeV $\to$ 1.71~GeV (red triangles). The photocouplings were taken from the RPP~\cite{rpp} (blue filled triangles) and the CLAS data analysis~\cite{Dug09} of $\pi N$ photoproduction (blue filled squares). }
\label{p11d13s31}
\end{figure}

Table~\ref{tab-2} summarizes the available CLAS results on the $\gamma_vpN^*$ electrocouplings. The resonance electrocouplings have been obtained from various CLAS data sets in the exclusive channels: $\pi^+n$ and $\pi^0p$ at $Q^2 < 5.0$~GeV$^2$ in the mass range up to 1.7~GeV, $\eta p$ at $Q^2 < 4.0$~GeV$^2$ in the mass range up to 1.6~GeV, and $\pi^+\pi^-p$ at $Q^2 < 1.5$~GeV$^2$ in the mass range up to 1.8~GeV~\cite{Bu12,Az09,park15,Mo12,Mo16}. The numerical values of these electrocouplings from CLAS can be found in Ref.~\cite{mokeev-web}. The computer code for interpolation/extrapolation over $Q^2$ in the range of $Q^2$ up to 5.0 GeV$^2$ of the CLAS results on resonance electrocouplings is available on the web page of Ref.~\cite{isupov-web}. Consistent results for the $\gamma_vpN^*$ electrocouplings of the $N(1440)1/2^+$ and $N(1520)3/2^-$ resonances, which have been determined in independent analyses of the dominant meson electroproduction channels $\pi N$ and $\pi^+\pi^-p$, shown in Fig.~\ref{p11d13s31} (left) and (middle), demonstrate that the extraction of these fundamental quantities is reliable. Studies of the exclusive electroproduction channels off protons $\pi N$ and $\pi^+\pi^-p$ offer complementary information on the $N^*$ electrocouplings. For low lying excited nucleon states in the mass range up to 1.6~GeV that decay preferentially to the $\pi N$ final states, the data on single-pion exclusive electroproduction drive the extraction of these resonance electrocouplings. However, studies the $\eta p$ and $\pi^+\pi^-p$ channels are needed in order to validate the $\gamma_vpN^*$ electrocoupling extraction from $\pi N$ electroproduction. 

The CLAS data for the $\pi^+ \pi^- p$ channel play a critical role in the extraction of the $\gamma_vpN^*$ electrocouplings of  higher-lying nucleon excited states ($M >1.60$~GeV), which decay preferentially to the $\pi\pi N$ final states, e.g. $\Delta(1620)1/2^-$,  $\Delta(1700)3/2^-$, $N(1720)3/2^+$, and the $N'(1720)3/2^+$ candidate state. Right now, the electrocouplings of these states can only be determined from the data in the $\pi^+\pi^-p$ exclusive electroproduction channel off protons, while the $\pi N$ channels do not have enough sensitivity to the electrocouplings of the aforementioned resonances. 

We have developed special procedures to test the reliability of the $\gamma_vpN^*$ resonance electrocouplings extracted from the charged double pion electroproduction data. In this case, we carried out the extraction of the resonance parameters, independently fitting the CLAS $\pi^+\pi^-p$ electroproduction data~\cite{Ri03} in overlapping intervals of $W$. The non-resonant amplitudes in each of the $W$-intervals are different, while the resonance parameters should remain the same as 
they are determined from the data fit in different $W$-intervals, see Fig.~\ref{p11d13s31} (right). The consistent results on 
these electrocouplings from the independent analyses in different $W$-intervals strongly support their reliable extraction. The tests described above demonstrated the capability of the models   to provide reliable information on the $\gamma_vpN^*$ resonance electrocouplings from independent analyses of the data on exclusive $\pi N$ and $\pi^+\pi^-p$ electroproduction. 

Exclusive $KY$ ($K\Lambda$, $K\Sigma$) electroproduction channels offer an independent source of the information on electrocouplings of resonances in the mass range of $M > 1.6$~GeV. Moreover, the $KY$ exclusive photoproduction channels have demonstrated particular sensitivity to the contributions from the new so-called ``missing" baryon states \cite{Bu17n,BnGa12,BnGa14}. The current status in resonance electroexcitation studies on strangeness electroproduction data with CLAS was reviewed in another contribution to these Proceedings~\cite{Car17n}. CLAS has provided sufficient experimental information for the extraction of the $\gamma_vpN^*$ resonance electrocouplings from $KY$ exclusive electroproduction data off protons, but the reaction models capable of describing these data with the quality needed to extract the resonance parameters are still not available. The development of reaction models capable of determining resonance electrocouplings from exclusive $KY$ electroproduction data is urgently needed to foster the exploration of high-lying nucleon resonances in different exclusive meson electroproduction channels, as well as for the search for new baryon states in combined studies of exclusive photo- and electroproduction data. 

\begin{figure}[htb!]
\centering
\includegraphics[width=5.4cm,clip]{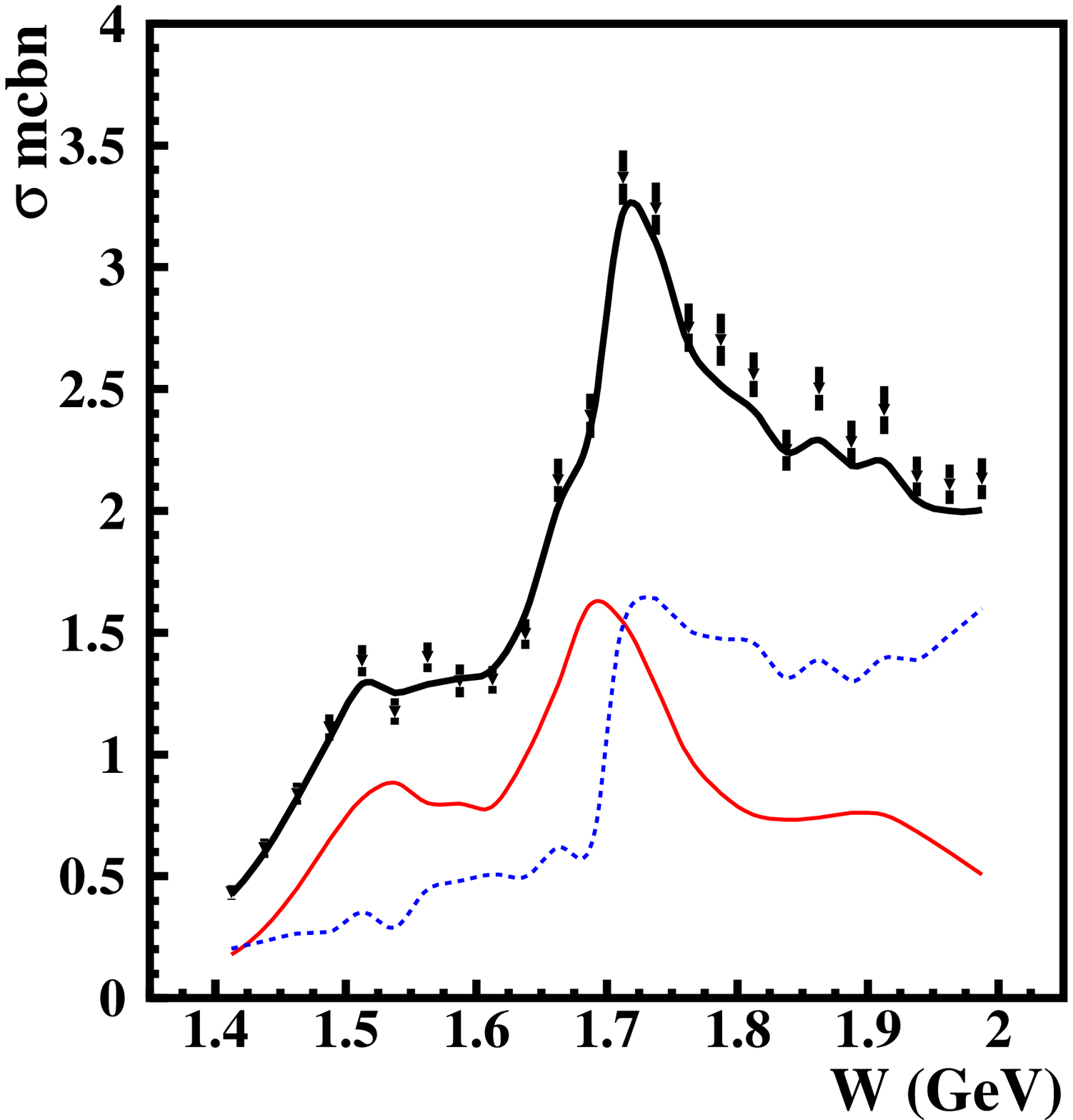}
\includegraphics[width=6.2cm,clip]{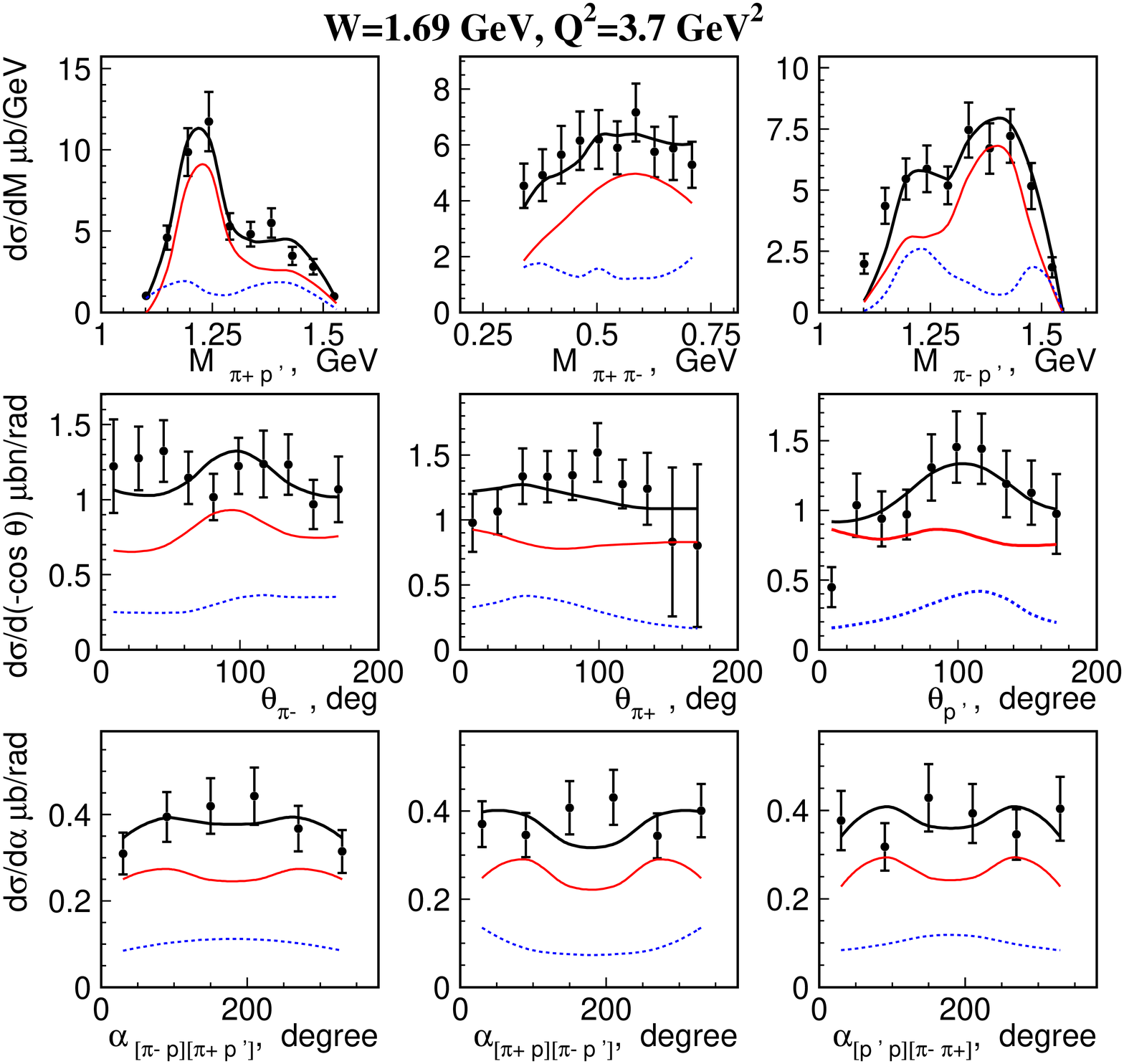}
\caption{(Color Online) Description of the experimental data on $\pi^+\pi^-p$ electroproduction off protons at 
1.4~GeV $< W < 2.0$~GeV and 2.0 GeV$^2 < Q^2 < 5.0$~GeV$^2$~\cite{Is17} achieved within the framework of the JM model. (Left) Preliminary results on the description of the fully integrated cross section at 3.5 GeV$^2 < Q^2 < 4.0$~GeV$^2$. (Right) Representative example for the description of the nine one-fold differential cross sections at $W$=1.69~GeV in the 
same $Q^2$-interval.  Thick black solid, red solid, and blue dashed curves show the computed full, resonant, and non-resonant contributions, respectively. Only the statistical uncertainties of the data are shown in the plots. } 
\label{twopihighq2}
\end{figure}

The CLAS Collaboration keeps gradually extending the kinematic coverage of the experimental data on $\pi^+n$, $\pi^0p$, and $\pi^+\pi^-p$ electroproduction off protons over $W$ and $Q^2$  with the goal to determine electrocouplings of high-lying nucleon resonances ($M > 1.6$~ GeV) in a wide range of photon virtualities up to 5.0 GeV$^2$. New data on $\pi^+\pi^-p$ exclusive meson electroproduction off protons at 1.4 GeV $< W <$ 2.0~GeV and 2.0 GeV$^2$ $< Q^2 <$ 5.0~GeV$^2$ have been recently published~\cite{Is17}. A good description of these data was achieved within the framework of the JM reaction model with $\chi^2/d.p.$ $<$ 1.6. The analysis of these data within the JM model demonstrated that the resonance contributions to all nine one-fold differential cross sections have distinctively different shapes in comparison with the non-resonant contributions and the relative resonance contribution increases with $Q^2$ (see Fig.~\ref{twopihighq2}), offering promising prospects for the extraction of the resonance electrocouplings. In the near term future, we are expecting to obtain $\gamma_vpN^*$ electrocouplings for most excited nucleon states in the mass range up to 2.0~GeV and $Q^2 < 5.0$~GeV$^2$ from independent studies on $\pi N$ and $\pi^+\pi^-p$ exclusive electroproduction off protons. 

\boldmath
\section{Resonance electrocouplings as a window into excited nucleon structure and strong QCD dynamics}
\unboldmath

Experimental results on the $\gamma_vpN^*$ electrocouplings offer insight into excited nucleon structure and the 
dynamics of strong QCD. Studies of the strong QCD dynamics that generates the structure of resonances require 
multi-prong approaches. These include continuum DSE and lattice QCD studies of the resonance structure starting from the QCD Lagrangian, as well as different quark models capable of describing the structure of many excited nucleon states, but by employing phenomenological parameterization for the $N^*$ generation mechanisms without a close connection to the QCD-Lagrangian.

\begin{figure}[htb!]
\centering
\includegraphics[width=5.8cm,clip]{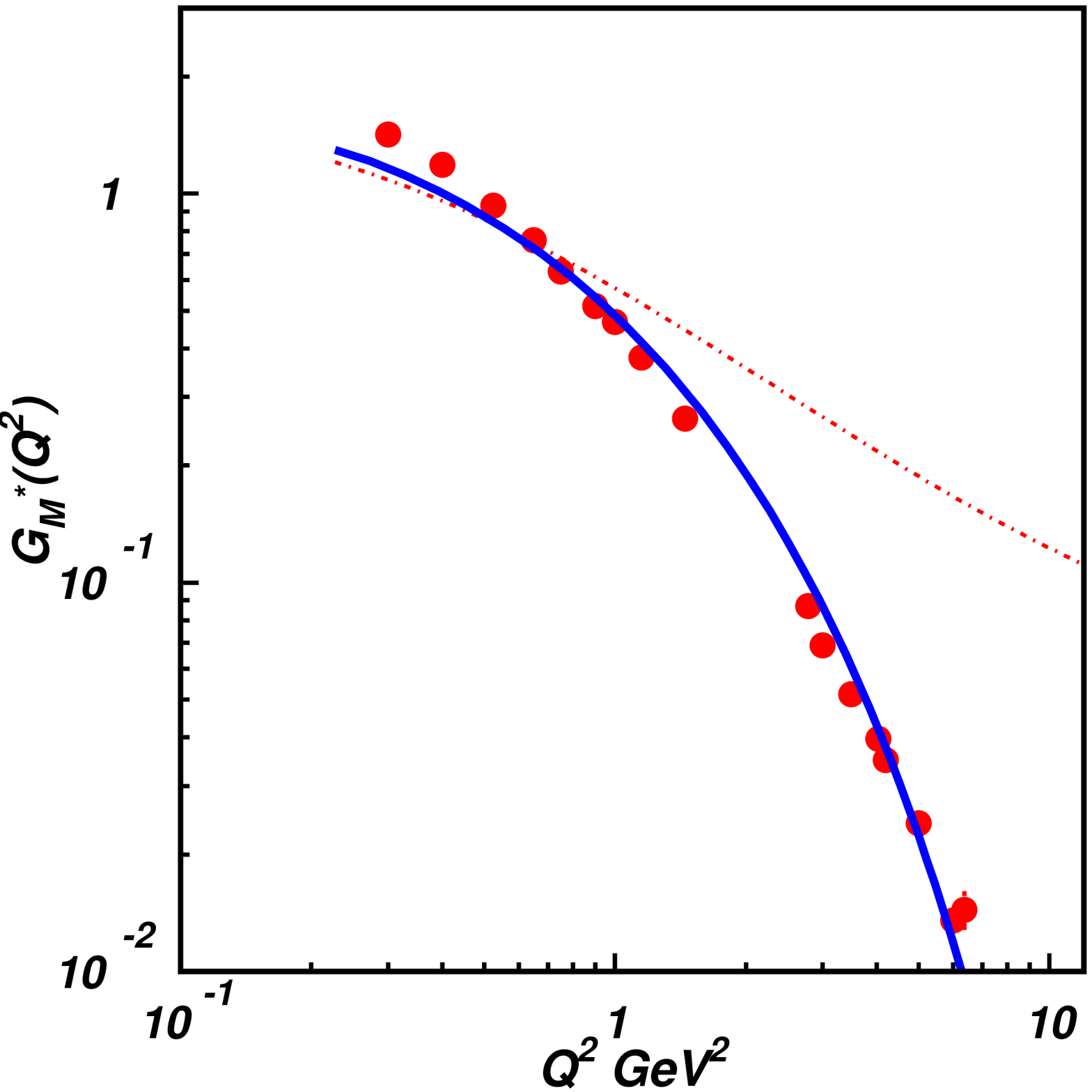}
\includegraphics[width=5.2cm,clip]{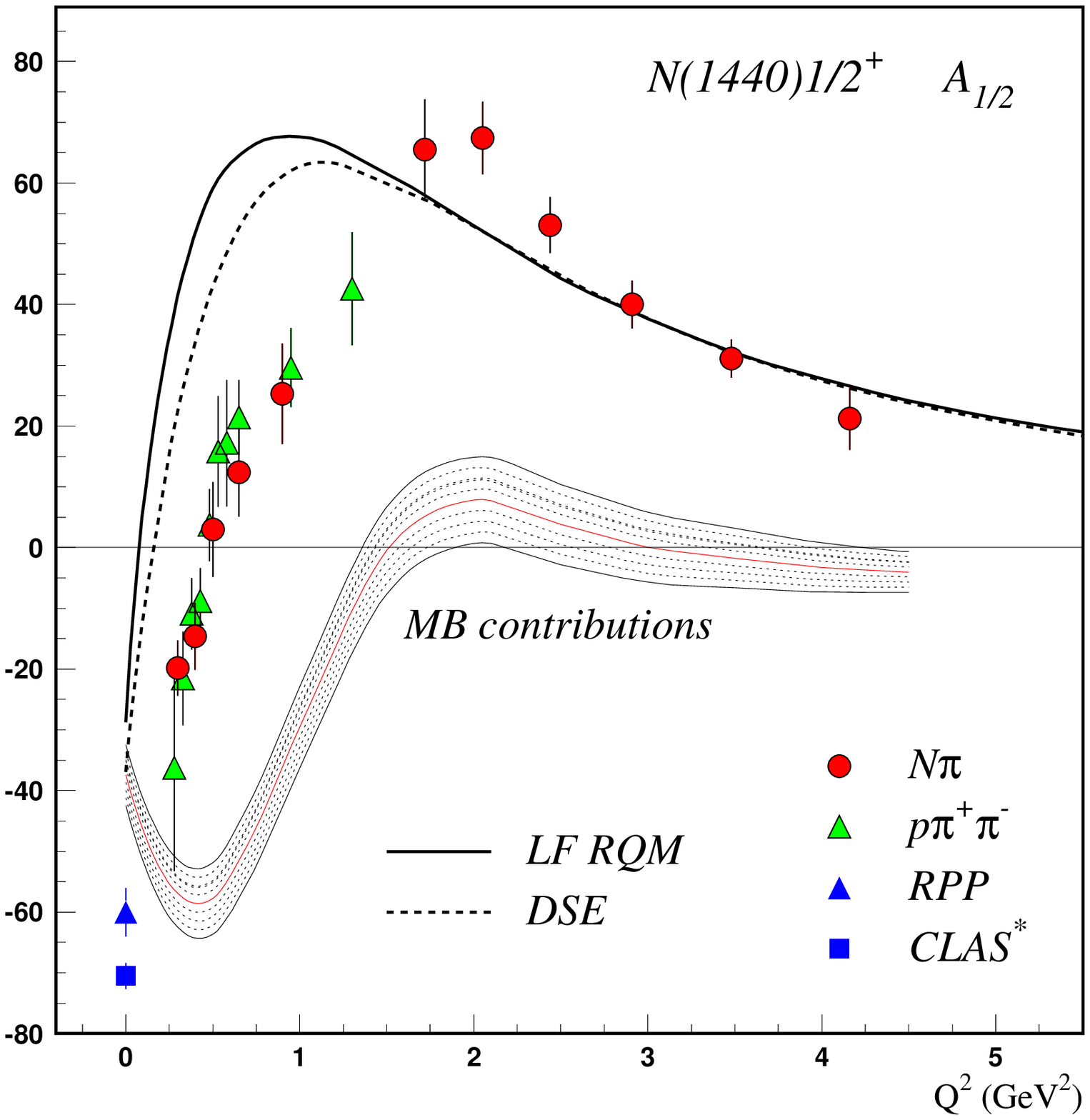}
\caption{(Color Online) Insight into the resonance structure and strong QCD dynamics from the experimental results on the 
magnetic $p \to \Delta(1232)3/2^+$ transition form factor (left) \cite{Az09} and the $A_{1/2}$ electrocoupling of the $N(1440)1/2^+$ resonance (right)~\cite{Az09,Mo12,Mo16}. The red dashed and blue solid curves in the left panel correspond 
to the computation of the magnetic $p \to \Delta(1232)3/2^+$ transition form factor starting from the QCD-Lagrangian within DSE by employing a simplified contact $qq$-interaction (frozen quark mass) and a realistic $qq$-interaction (running quark mass), respectively \cite{Seg14}. In the right panel the DSE computation of the $A_{1/2}$ electrocoupling of the $N(1440)1/2^+$ resonance with a realistic $qq$-interaction and the same running quark mass as employed in the successful evaluations of nucleon elastic and magnetic $p \to \Delta(1232)3/2^+$ form factors is shown by the dashed line~\cite{Seg15}. The result from a novel light front quark model \cite{Az12,Az17}, which incorporates a momentum dependent dressed quark mass, is shown by the solid line. The shadowed area represents the meson-baryon cloud contribution inferred from the experimental results on the $A_{1/2}$ electrocoupling of the $N(1440)1/2^+$ resonance and the DSE evaluation \cite{Seg15} of the quark core contribution.}
\label{del33p11}
\end{figure}

Due to the rapid progress in the field of DSE studies of excited nucleon states~\cite{Ro17n,Ro17,Seg14,Seg15}, the first evaluations of the magnetic $p \to \Delta(1232)3/2^+$ form factors and the $N(1440)1/2^+$ resonance electrocouplings 
starting from the QCD Lagrangian have recently become available. The $p \to \Delta(1232)3/2^+$ magnetic form factor and $A_{1/2}$ electrocoupling of the $N(1440)1/2^+$ resonance computed in Refs.~\cite{Seg14,Seg15} are shown in Fig.~\ref{del33p11}. These evaluations are applicable at photon virtualities  where the contributions of the inner quark core to the resonance electrocouplings are much larger than those from the external meson baryon cloud. In this range of photon 
virtualities, the evaluations~\cite{Seg14,Seg15} offer a good description of the experimental results on the 
$p \to \Delta(1232)3/2^+$ transition form factors and the $N(1440)1/2^+$ resonance electrocouplings.  

Analysis of the CLAS results \cite{Az09}  on the magnetic $p \to \Delta(1232)3/2^+$  form factor within DSE demonstrated for the first time that the masses of dressed quarks are in fact running with quark momentum as predicted by the DSE computations of the dressed quark mass function starting from the QCD-Lagrangian. The DSE evaluation of the magnetic $p \to \Delta(1232)3/2^+$ form factor was carried out by employing the simplified contact qq-interaction (dashed lines in Fig.~\ref{del33p11} (left)) and with the most advanced realistic $qq$-interaction~\cite{Bi17,Ch18} (solid lines in Fig.~\ref{del33p11} (left)). The contact $qq$-interaction produces a dynamically generated dressed quark mass of 
$\approx$400~MeV, that is momentum independent. DSE computations with a realistic $qq$-interaction~\cite{Bi17,Ch18} 
predict a momentum dependent quark mass as shown in Fig.~\ref{qmass}. The DSE results with a frozen quark mass overestimate the CLAS data at $Q^2 > 1.5$~GeV$^2$. The discrepancies are increasing with $Q^2$. Instead, by employing a quark mass that is running with momentum (with the parameterization shown in Fig.~\ref{qmass}), the DSE computations offer 
a good description of the experimental results on the $p \to \Delta(1232)3/2^+$ magnetic form factor in the entire range of photon virtualities 0.8 GeV$^2$ $<$ $Q^2$ $<$ 7.0 GeV$^2$.   

Remarkably, a good description of the experimental results on the $p \to \Delta(1232)3/2^+$ transition form factors and the $N(1440)1/2^+$ resonance electrocouplings is achieved with a momentum dependence of the dressed quark mass that is 
{\it exactly the same} as the one employed in the previous evaluations of the elastic electromagnetic nucleon form factors~\cite{Seg14}. This success strongly supports: a) the relevance of dynamical dressed quarks with properties predicted 
by the DSE approach~\cite{Ro17n,Bi17,Ch18}, as constituents of the quark core for the structure both of the ground and 
excited nucleon states and b) the capability of the DSE approach~\cite{Seg14,Seg15} to map out the dressed quark mass 
function from the experimental results on the $Q^2$-evolution of the nucleon elastic and $p \to N^*$ electromagnetic transition form factors, or rather $\gamma_vpN^*$ electrocouplings.

\begin{figure}[htb!]
\centering
\includegraphics[width=5.7cm,clip]{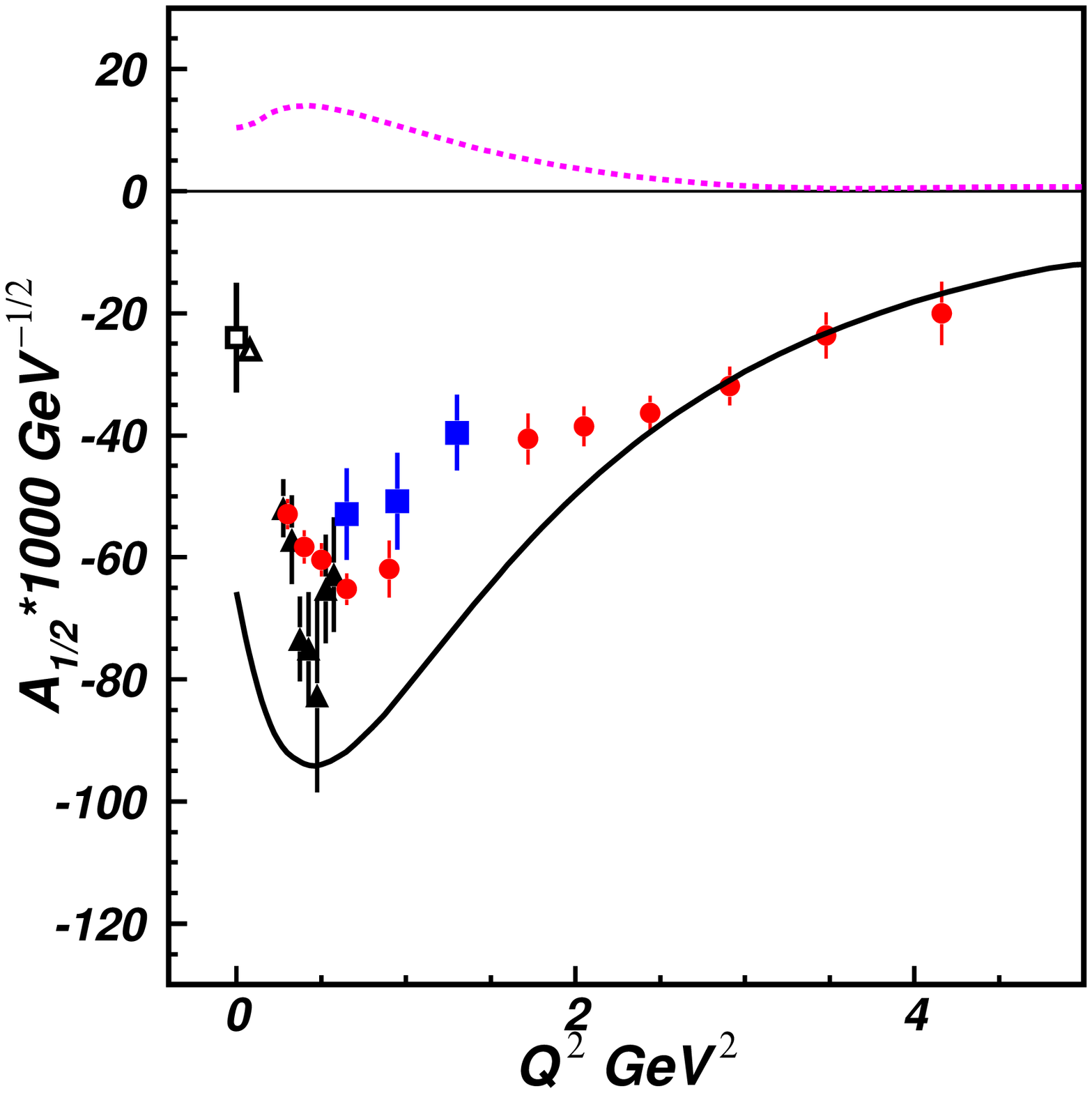}
\includegraphics[width=5.8cm,clip]{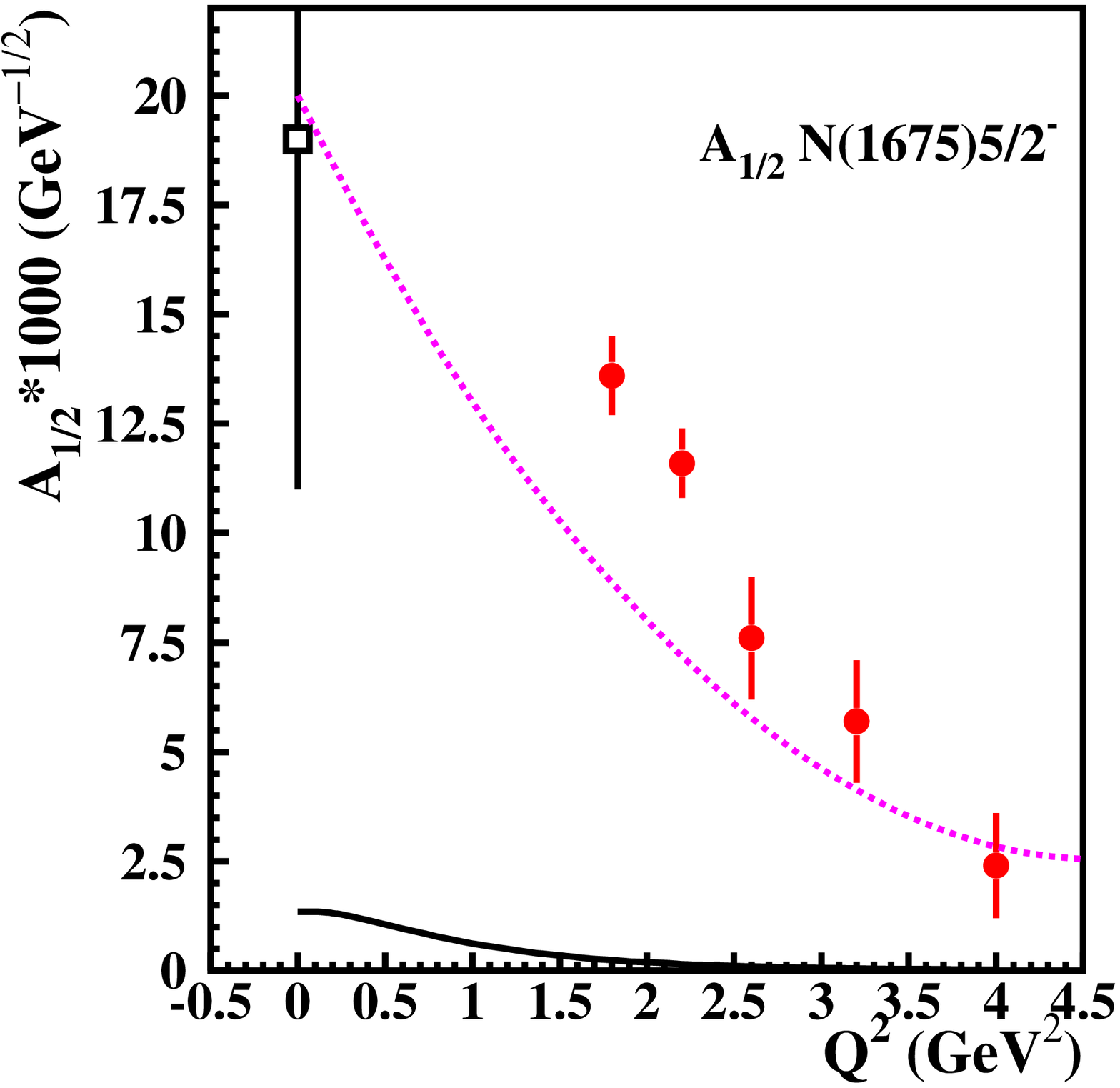}
\caption{(Color Online) Contributions from the meson-baryon cloud (dashed magenta) and the quark core (solid black) to the $A_{1/2}$ electrocouplings of the $N(1520)3/2^-$ (left) and $N(1675)5/2^+$ resonances evaluated within the framework of the Argonne-Osaka approach \cite{Lee08} (absolute values) for the meson-baryon cloud and the hQCM-model~\cite{San12} for the quark core.}
\label{d13d15}
\end{figure}

The $\gamma_vpN^*$ electrocouplings of many excited nucleon states in the mass range up to 1.7~GeV were evaluated within a novel light front quark model (LFQM)~\cite{Az12,Az17}. This model accounts for the contributions from both the meson-baryon cloud and the quark core, and incorporates the parameterized momentum dependent quark mass with parameters adjusted to the data on the $Q^2$-evolution of the nucleon elastic form factors. It was found that the implementation of the momentum dependent dressed quark mass is absolutely needed in order to reproduce the behavior of the nucleon elastic form factors at $Q^2$ $>$ 2.0 GeV$^2$. A successful description of the electrocouplings of most resonances in the mass range up to 1.7~GeV was achieved with the same momentum dependent quark mass used for the successful description of the nucleon elastic form factors. A typical example for the description of the $A_{1/2}$ $N(1440)1/2^+$ electrocouplings is shown in Fig.~\ref{del33p11} (right) by the solid line. A successful description of the $\gamma_vpN^*$ electrocouplings for most excited nucleon states in the mass range up to 1.7~GeV offers support for the running dressed quark mass from a framework conceptually different than the DSE approach. 

The analysis of the CLAS results on the $\gamma_vpN^*$ electrocouplings of most excited nucleon states in the mass range 
up to 1.7~GeV has revealed the $N^*$ structure for $Q^2 < 5.0$~GeV$^2$ as a complex interplay between an inner core of three dressed quarks and an external meson-baryon cloud~\cite{Bu17n,Bu12,Ka17,Az12,Az17,Seg15}. The credible DSE evaluation of the quark core contributions to the electrocouplings of the $N(1440)1/2^+$ state~\cite{Seg15} has allowed us 
to infer the meson-baryon cloud contributions to this resonance as the difference between the experimental data on the resonance electrocouplings and the quark core electroexcitation amplitudes computed from DSE, as shown by the shadowed area in Fig.~\ref{del33p11} (right). The relative contributions of the quark core and the meson-baryon cloud depend strongly on the quantum numbers of the excited nucleon state. The quark core becomes the dominant contributor to the $A_{1/2}$ electrocouplings of the $N(1440)1/2^-$ and $N(1520)3/2^-$ resonances at $Q^2 > 2.0$~GeV$^2$, as can be seen 
in Fig.~\ref{del33p11} (right) and Fig.~\ref{d13d15} (left), respectively. These electrocouplings offer almost direct access to the dressed quark contributions for $Q^2 > 2.0$~GeV$^2$. Instead, the electrocouplings of the $N(1675)5/2^-$ state, shown in 
Fig.~\ref{d13d15} (right), are dominated by meson-baryon cloud contributions, which allows us to explore this component from 
the electrocoupling data. The relative contributions of the meson-baryon cloud to the electrocouplings of all resonances studied with CLAS decrease with $Q^2$ in a gradual transition towards quark core dominance at photon virtualities above 5.0~GeV$^2$. 

\boldmath
\section{Future studies of nucleon resonances with the CLAS12 detector}
\label{clas12-program}
\unboldmath

After completion of the Jefferson Lab 12~GeV Upgrade Project, the commissioning run for the CLAS12 detector in the upgraded 
Hall~B started successfully at the end of 2017`\cite{clas12}. CLAS12 will be the only foreseable facility worldwide capable of studying nucleon resonances in the still unexplored ranges of the smallest photon virtualities 0.05~GeV$^2 < Q^2 < 0.5$~GeV$^2$ and the highest photon virtualities up to 12~GeV$^2$ ever achieved in exclusive reaction measurements~\cite{Bu17n,Ro17n,Az13}.

The studies of nucleon resonances at small photon virtualities are driven by the search for new states of baryon matter, the so-called hybrid-baryons~\cite{e12-16-010}. Small $Q^2$ is preferential for the observation of these new states that contain three dressed quarks and, in addition, glue as a structural component. The LQCD studies of the $N^*$ spectrum starting from the QCD Lagrangian~\cite{Du12} predict several such states in the mass range from 2.0~GeV to 2.5~GeV after reducing the predicted hybrid mass values by the differences between the experimental results on the masses of the known lightest $N^*$ 
of the same spin-parities as for the expected hybrid baryons and their values from LQCD. 

In the experiment with CLAS12, we will search for the hybrid signal as the presence of extra states in the conventional resonance spectrum of $J^P$=1/2$^+$, 3/2$^+$ in the mass range from 2.0~GeV to 2.5~GeV from the data on exclusive 
$KY$ and $\pi^+\pi^-p$ electroproduction off protons~\cite{e12-16-010}. The hybrid nature of the new baryon 
states will be identified by looking for the specific $Q^2$ evolution of their electrocouplings. We expect a specific behavior of the hybrid state electrocouplings with $Q^2$ because one might imagine that the three quarks in a hybrid baryon should be in a color-octet state in order to create a colorless hadron in combination with the glue constituent in a color-octet state.  Instead, in regular baryons, dressed quarks should be in a color-singlet state, so pronounced differences for quark configurations in the structure of conventional and hybrid baryons should results in a peculiar $Q^2$ evolution of hybrid baryon electrocouplings. The sound theoretical predictions for the evolution of hybrid-resonance electrocouplings with photon virtualities are critical for hybrid-baryon identification. They represent an urgently needed part of the theory support for the hybrid-baryon search with CLAS12. 

The studies on $N^*$ structure at low $Q^2$ over the spectrum of all prominent resonances will also complete the long term efforts on the search for the new ``missing" baryon states. The successful description of the combined photo- and electroproduction data in the 
entire range of $Q^2$ by employing a $Q^2$ independent resonance mass, as well as $Q^2$ independent total and partial hadronic decay widths, will validate the presence of new states in the baryon spectrum in a nearly model independent way. The extension of amplitude analysis~\cite{BnGa12,BnGa14,Asz16,Str16} and coupled channel analysis~\cite{Ka13} methods that were successfully employed in the studies of exclusive meson photoproduction to the electroproduction off protons is of particular importance for the success of the aforementioned efforts.

Exploration of the excited nucleon state structure in exclusive $\pi N$, $KY$, and $\pi^+\pi^-p$ electroproduction off protons at 5.0~GeV$^2 < Q^2 < 12.0$~GeV$^2$~\cite{e12-09-003,e12-06-108a,e12-16-010a} is scheduled in the first year of running with the CLAS12 detector. For the first time, the electrocouplings of all prominent nucleon resonances will become available at the highest photon virtualities ever achieved in the studies of exclusive electroproduction. These distance scales correspond to the still unexplored regime for $N^*$ electroexcitations where the resonance structure is dominated by the quark core with almost negligible meson-baryon cloud contributions. The foreseen experiments offer almost direct access to the properties of dressed quarks inside $N^*$ states of different quantum numbers. Consistent results on the dressed quark mass function derived from independent analyses of the data on the $\gamma_vpN^*$ electrocouplings of the resonances with distinctively different structure, such as radial excitations, spin-isospin flip, and orbital excitations, will validate the credible access to this fundamental ingredient of strong QCD from the experimental data. The expected data on the $\gamma_vpN^*$ electrocouplings will provide for the first time access to the dressed quark mass function in the range of quark momenta up to 1.5~GeV, where the transition from the quark-gluon confinement to the pQCD regimes of the strong interaction takes full effect, as is shown in Fig.~\ref{qmass}. Exploring 
the dressed quark mass function at these distances will allow us to address the most challenging open problems 
of the Standard Model on the nature of $>$98\% of hadron mass and quark-gluon confinement~\cite{Ro17n,Az13}. 

The future prospects for the evaluation of the nucleon resonance electrocouplings within lattice QCD~\cite{Br17n} suggest the possibility to employ the CLAS12 results for the exploration of the baryon structure emergence from first principles of QCD. Quark models will remain vital for the exploration of the resonance structure  over the full spectrum of excited nucleon states accessible with the CLAS12 data.

The success of the program on nucleon resonance studies requires synergistic efforts between experimentalists, reaction 
model phenomenologists, and hadron structure theorists. Studies of the nucleon resonance spectrum and structure represent 
a direction of key importance in the broad worldwide efforts focused on the exploration of strong QCD dynamics. Indeed, as remarked at the turn of this century~\cite{Isgur2000}: "[baryons and their resonances] must be at the center of any discussion 
of why the world we actually experience has the character it does."  This is a primary motivation for our efforts.
  
\begin{acknowledgements} 
This work was supported by the U.S. Department of Energy, Office of Science, Office of Nuclear Physics under contract DE-AC05-06OR23177. The author is grateful for many stimulating and fruitful discussions on different topics of this paper with I.G. Aznauryan,  V.D. Burkert, D.S. Carman, R.W. Gothe, K. Hicks, K. Joo, T-S. H. Lee, and C.D. Roberts. 
\end{acknowledgements}

\end{document}